\documentclass[traditabstract]{aa}
\usepackage{txfonts}
\usepackage{graphicx}
\usepackage{natbib}
\bibpunct{(}{)}{;}{a}{}{,}

\begin{document}

\title{Parameters of two low-mass contact eclipsing binaries near the short-period limit}
\author{M.~E.~Lohr\inst{\ref{inst1}}\and S.~T.~Hodgkin\inst{\ref{inst2}}\and
  A.~J.~Norton\inst{\ref{inst1}}\and U.~C.~Kolb\inst{\ref{inst1}}}
\institute{Department of Physical Sciences, The Open University,
  Walton Hall, Milton Keynes MK7\,6AA, UK\\ \email{Marcus.Lohr@open.ac.uk}\label{inst1}\and Institute of Astronomy, Madingley Road, Cambridge CB3\,0HA, UK\label{inst2}}
\date{Received 21 May 2013 / Accepted 31 January 2014}

\abstract {The two objects \object{1SWASP~J150822.80$-$054236.9} and
  \object{1SWASP~J160156.04+202821.6} were initially detected from
  their SuperWASP archived light curves as candidate eclipsing
  binaries with periods close to the short-period cut-off of the
  orbital period distribution of main sequence binaries, at
  $\sim$0.2~d.  Here, using INT spectroscopic data, we confirm them as
  double-lined spectroscopic and eclipsing binaries, in contact
  configuration.  Following modelling of their visual light curves and
  radial velocity curves, we determine their component and system
  parameters to precisions between $\sim$2 and 11\%.  The former
  system contains 1.07 and 0.55~$M_{\sun}$ components, with radii of
  0.90 and 0.68~$R_{\sun}$ respectively; its primary exhibits
  pulsations with period 1/6 the orbital period of the system.  The
  latter contains 0.86 and 0.57~$M_{\sun}$ components, with radii of
  0.75 and 0.63~$R_{\sun}$ respectively.}

\titlerunning{Parameters of two eclipsing binaries near the
  short-period limit}
\authorrunning{M.~E.~Lohr et al.}

\maketitle
\section{Introduction}

\begin{figure}
\resizebox{\hsize}{!}{\includegraphics{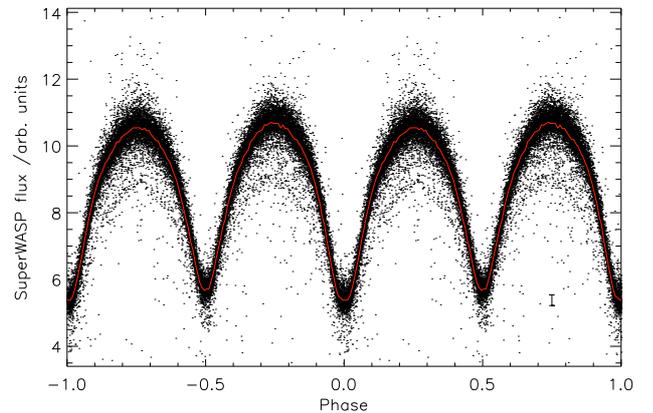}}
\caption{SuperWASP light curve for J150822, folded at period of
  22469.219~s, with binned mean curve overplotted.  A typical
    uncertainty range for a single observation is shown.  These fluxes
    correspond to a visual magnitude range of $\sim$12.4--13.2.}
\label{star10lc}
\end{figure}

\begin{figure}
\resizebox{\hsize}{!}{\includegraphics{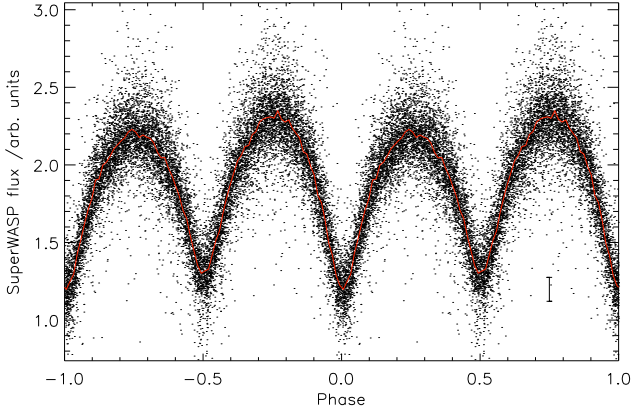}}
\caption{SuperWASP light curve for J160156, folded at period of
  19572.136~s, with binned mean curve overplotted.  A typical
    uncertainty range for a single observation is shown.  These fluxes
    correspond to a visual magnitude range of $\sim$14.1--14.8.}
\label{star28lc}
\end{figure}

The orbital period distribution of main sequence binary stars exhibits
a fairly sharp lower limit at around 0.2~d \citep{ruc92, ruc07,
szymanski, paczynski}, the cause of which is the subject of ongoing
research e.g. \citet{stepien, stepien12, jiang}.  However, despite the
inherent interest of this region of parameter space, relatively few
eclipsing binaries (EB) have been discovered with periods near the
cut-off point.  This motivated a search of the archive of the
SuperWASP project (Wide Angle Search for Planets: \citet{pollacco})
for EB candidates with apparent periods $<$20\,000~s ($\sim$0.2315~d),
reported in \citet{norton}.  53 candidates were found, 48 of which
were new discoveries at the time.  A subsequent search of these
candidate EBs for evidence of period change \citep{lohr} corrected the
periods of nine to values slightly greater than 20\,000~s, but still
$<$22600~s ($\sim$0.2616~d).  A more rigorous search of the SuperWASP
archive \citep{lohr13} then detected 143 candidate EBs with periods
$<$20\,000~s, including 97 new discoveries since \citet{norton}, and
measured significant period changes in 74 candidates.

Here, spectroscopic data allow us to confirm two of these candidates
as double-lined EBs in contact configuration (W\,UMa-type variables):
\object{1SWASP~J150822.80$-$054236.9} (J150822) and
\object{1SWASP~J160156.04+202821.6} (J160156).  Both were initially
identified in \citet{norton}; the period of J150822 was revised
upwards to 22469.2~s in \citet{lohr} and so it did not appear in
\citet{lohr13}.  We report system and component parameters obtained
for these EBs by simultaneous modelling of their SuperWASP light
curves and radial velocities.  These should be of interest for the
study of low-mass dwarfs and W\,UMa systems in general, and of very
short period binaries specifically.

\section{Observations}

\subsection{Photometry}

\begin{figure}
\resizebox{\hsize}{!}{\includegraphics{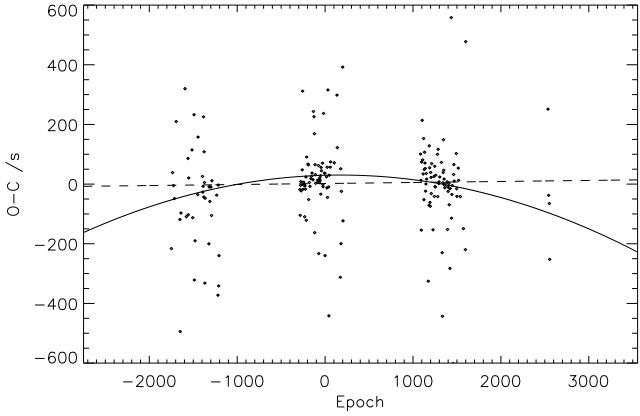}}
\caption{Observed minus calculated (O$-$C) diagram for J150822, with
best linear (dashed line; $\chi^2=5.25$) and quadratic (solid curve;
$\chi^2=4.87$) fits overplotted.  For clarity of presentation,
uncertainties are not shown.  Period decrease of
$-0.055\pm0.006$~s~y\textsuperscript{-1} is indicated ($\sigma=8$).}
\label{star10oc}
\end{figure}

\begin{figure}
\resizebox{\hsize}{!}{\includegraphics{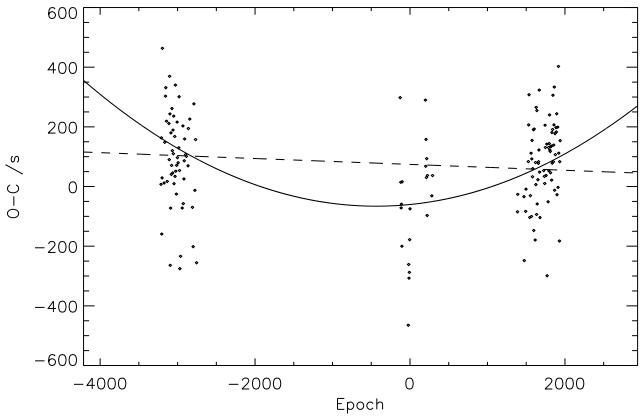}}
\caption{O$-$C diagram for J160156, with best linear (dashed line;
$\chi^2=1.87$) and quadratic (solid curve; $\chi^2=1.64$) fits
overplotted.  For clarity of presentation, uncertainties are not
shown.  Period increase of $+0.094\pm0.015$~s~y\textsuperscript{-1} is
indicated ($\sigma=6$).}
\label{star28oc}
\end{figure}

The SuperWASP archive contains 30131 photometric points for J150822,
taken between 5 March 2008 and 29 March 2011.  For J160156 there are
14651 observations, made between 2 May 2004 and 21 February 2011.
Sys-Rem-corrected fluxes \citep{tamuz, mazeh} from the
3.5~pixel-radius photometric aperture (the middle of three available
apertures) were used to construct the light curves used here, which
correspond approximately to the Johnson V band.  Periods and
period-change measurements were obtained using a custom-written IDL
program, as described in \citet{lohr13}, and the binned averaged
phase-folded data produced high-precision phased light curves
(Figs.~\ref{star10lc} to \ref{star28oc}).  A small secular
period decrease ($-$0.055~s~y\textsuperscript{-1}) was measured for
J150822, and a slightly larger secular period increase
(0.094~s~y\textsuperscript{-1}) for J160156; both values are fairly
unexceptional for variables of this type \citep{lohr13}.  The scatter
here is comparable to that seen in other SuperWASP light curves for
objects of similar magnitude, and we do not believe that either period
variation over time, or flux variability caused by surface spots,
contribute to it significantly.  The uncertainties on the means in the
light curves used in subsequent modelling were given by the standard
deviation of points in each bin, divided by the square root of the
number of observations per bin i.e. $\sigma /\sqrt{n}$.

\subsection{Spectroscopy}

\begin{figure}
\resizebox{\hsize}{!}{\includegraphics{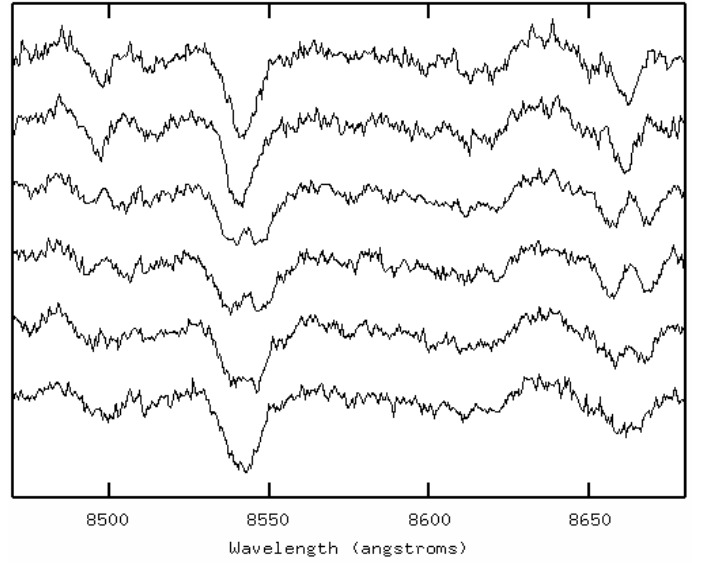}}
\caption{Selected spectra in region of \ion{Ca}{II} triplet
  (laboratory wavelengths: 8498.03, 8542.09 and 8662.14~$\AA$) for
  J150822, taken from final night of observations.  Line splitting is
  readily apparent for all three calcium lines.}
\label{linesplit10}
\end{figure}

\begin{figure}
\resizebox{\hsize}{!}{\includegraphics{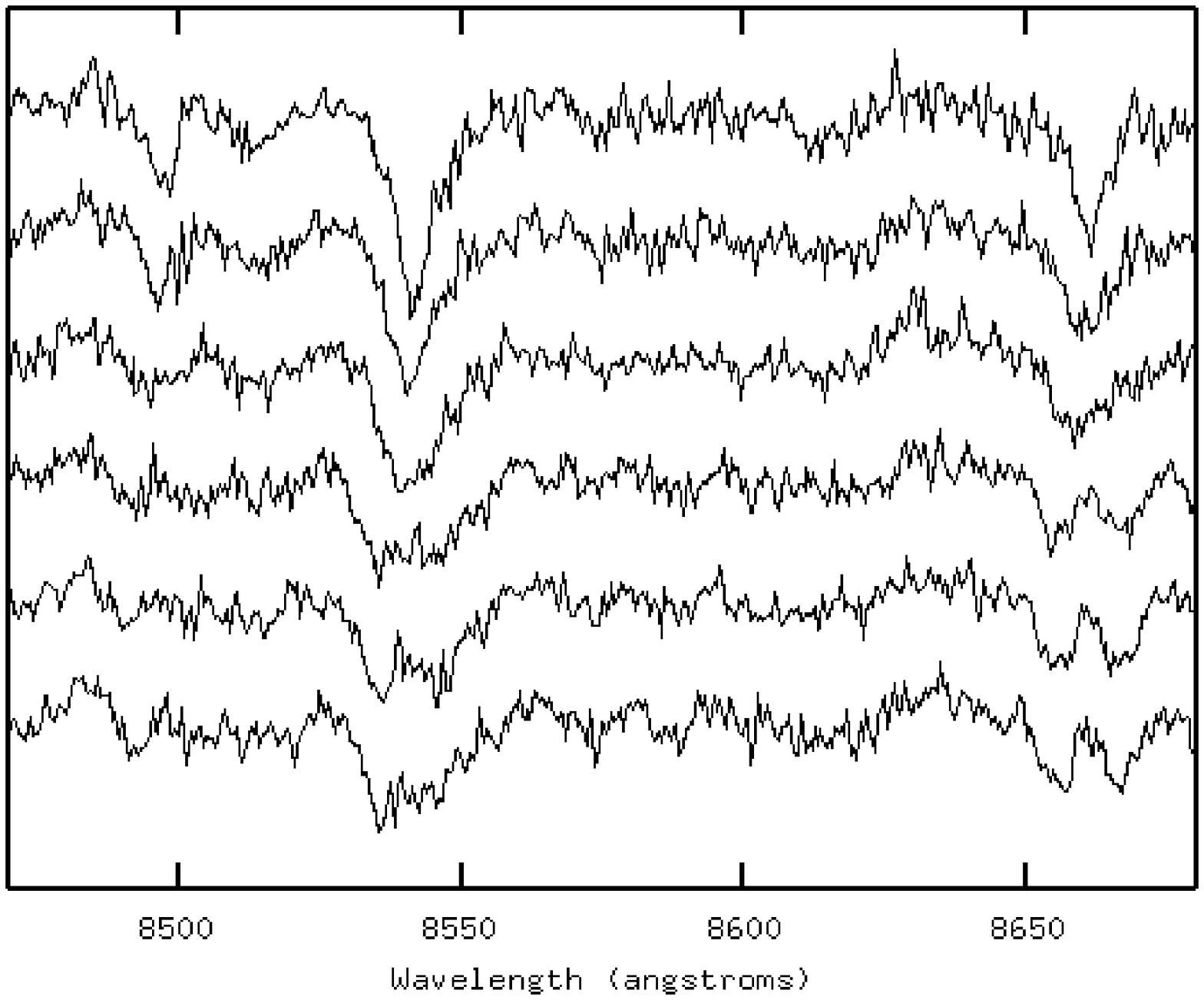}}
\caption{Selected spectra in region of \ion{Ca}{II} triplet for
J160156, taken from final night of observations.  Line splitting is
most obvious for the two calcium lines at longer wavelengths.}
\label{linesplit28}
\end{figure}

\begin{figure}
\resizebox{\hsize}{!}{\includegraphics{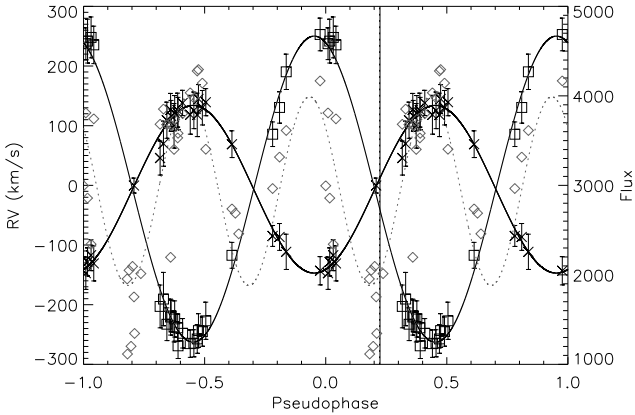}}
\caption{Radial velocities for J150822 (crosses indicate primary
component, squares secondary) with preliminary fits (solid curves)
used to obtain correct phasing of observations.  Also shown are a
light curve obtained from the spectra themselves (grey diamonds,
fitted with dotted grey curve) and the predictions for time of minimum
light from SuperWASP ephemerides (solid vertical line indicates linear
ephemeris i.e. no period change, dotted vertical line
quadratic i.e. secular period change; the two are almost
coincident).}
\label{star10phasedet}
\end{figure}

\begin{figure}
\resizebox{\hsize}{!}{\includegraphics{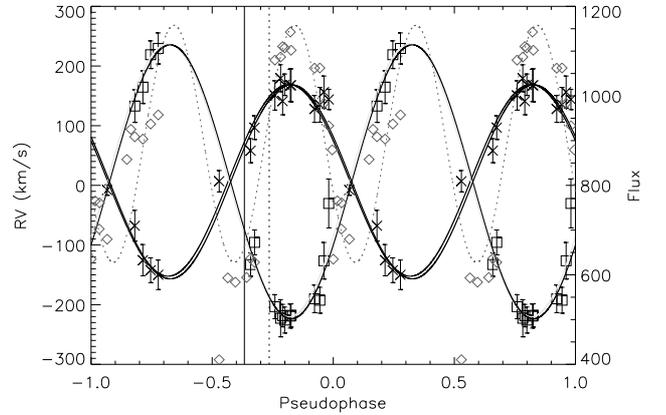}}
\caption{Radial velocities for J160156 (crosses indicate primary
component, squares secondary) with preliminary fits (solid curves)
used to obtain correct phasing of observations.  The primary and
secondary curve fits were determined consecutively, with the second
fit deriving some parameters from the first, and then refitted in the
other order; this has resulted in the visibly double fit curve for the
primary component.  Also shown are a light curve obtained from the
spectra themselves (grey diamonds, fitted with dotted grey curve) and
the predictions for time of minimum light from SuperWASP ephemerides
(solid vertical line indicates linear ephemeris, dotted vertical line
quadratic).}
\label{star28phasedet}
\end{figure}

\begin{table*}[!]
\caption{Summary of spectroscopic observations and derived quantities for J150822}
\label{star10table}
\centering
\begin{tabular}{l l l l l l l l l}
\hline\hline
HJD & Phase & Primary & $\delta$ Primary & Secondary & $\delta$
Secondary & Continuum flux at \\
$-2450000$ &  & RV (km~s\textsuperscript{-1}) & RV
(km~s\textsuperscript{-1}) & RV (km~s\textsuperscript{-1}) & RV
(km~s\textsuperscript{-1}) & 8500~$\AA$ (arb. units) \\
\hline
5997.5934 & 0.302 & \tablefootmark{a} &  &  &  & 2995 \\
5997.5977 & 0.319 & -128 & 24 & 242 & 37 & 2858 \\
5997.6015 & 0.333 & -117 & 26 & 248 & 34 & 2342 \\
5997.6406 & 0.484 &  &  &  &  & 1115 \\
5997.6444 & 0.498 &  &  &  &  & 1202 \\
5997.6482 & 0.513 &  &  &  &  & 1346 \\
5997.6756 & 0.618 & 46 & 29 & -203 & 36 & 3682 \\
5997.6792 & 0.632 & 70 & 38 & -191 & 45 & 3852 \\
5997.6829 & 0.646 & 109 & 30 & -233 & 27 & 3678 \\
5997.7080 & 0.743 & 118 & 33 & -265 & 23 & 4034 \\
5997.7119 & 0.758 & 126 & 31 & -265 & 23 & 3978 \\
5997.7155 & 0.772 & 118 & 36 & -257 & 24 & 4278 \\
5998.6826 & 0.490 &  &  &  &  & 2050 \\
5998.6863 & 0.504 &  &  &  &  & 2095 \\
5998.7272 & 0.661 & 127 & 28 & -221 & 30 & 2196 \\
5998.7309 & 0.676 & 100 & 37 & -237 & 25 & 3401 \\
5998.7348 & 0.691 & 126 & 28 & -239 & 23 & 3530 \\
5998.7571 & 0.776 & 145 & 24 & -243 & 28 & 4294 \\
5998.7613 & 0.793 & 130 & 29 & -245 & 24 & 4143 \\
5998.7650 & 0.807 & 139 & 23 & -227 & 31 & 3403 \\
5999.5733 & 0.915 & 69 & 22 & -117 & 22 & 2737 \\
5999.5770 & 0.929 &  &  &  &  & 2689 \\
5999.5806 & 0.943 &  &  &  &  & 2459 \\
5999.6170 & 0.083 & -85 & 18 & 86 & 21 & 2965 \\
5999.6244 & 0.111 & -86 & 23 & 131 & 26 & 3316 \\
5999.6316 & 0.139 & -111 & 30 & 190 & 30 & 3613 \\
5999.6680 & 0.279 & -143 & 24 & 253 & 27 & 4167 \\
5999.6765 & 0.312 & -147 & 27 & 237 & 27 & 3810 \\
5999.6850 & 0.344 & -131 & 29 & 235 & 31 & 3744 \\
5999.7209 & 0.482 &  &  &  &  & 1954 \\
5999.7282 & 0.510 & 0 & 13 &  &  & 1754 \\
5999.7356 & 0.539 &  &  &  &  & 2014 \\
5999.7681 & 0.664 & 116 & 26 & -219 & 27 & 3669 \\
5999.7719 & 0.678 & 121 & 29 & -224 & 30 & 3737 \\
5999.7755 & 0.692 & 121 & 27 & -269 & 21 & 3582 \\
5999.7804 & 0.711 & 138 & 21 & -248 & 25 & 3802 \\
\hline
\end{tabular}
\tablefoot{
\tablefoottext{a}{Radial velocities unusable.}
}
\end{table*}

\begin{table*}
\caption{Summary of spectroscopic observations and derived quantities for J160156}
\label{star28table}
\centering
\begin{tabular}{l l l l l l l l l}
\hline\hline
HJD & Phase & Primary & $\delta$ Primary & Secondary & $\delta$
Secondary & Continuum flux at \\
$-2450000$ &  & RV (km~s\textsuperscript{-1}) & RV
(km~s\textsuperscript{-1}) & RV (km~s\textsuperscript{-1}) & RV
(km~s\textsuperscript{-1}) & 8500~$\AA$ (arb. units) \\
\hline
5997.6567 & 0.923 &  &  &  &  & 635 \\
5997.6605 & 0.940 &  &  &  &  & 765 \\
5997.6643 & 0.957 &  &  &  &  & 702 \\
5997.6902 & 0.071 &  &  &  &  & 858 \\
5997.6939 & 0.087 &  &  &  &  & 925 \\
5997.6976 & 0.104 & -68 & 26 & 133 & 27 & 909 \\
5998.7083 & 0.565 &  &  &  &  & 594 \\
5998.7120 & 0.582 & 58 & 20 & -133 & 20 & 637 \\
5998.7157 & 0.598 & 97 & 20 & -96 & 21 & 628 \\
5998.7401 & 0.706 & 180 & 23 & -223 & 31 & 1087 \\
5998.7439 & 0.722 & 169 & 18 & -226 & 22 & 1107 \\
5998.7499 & 0.749 & 168 & 28 & -218 & 19 & 1102 \\
5998.7718 & 0.846 & 128 & 22 & -190 & 23 & 1062 \\
5998.7768 & 0.868 & 135 & 29 & -193 & 22 & 1062 \\
5998.7806 & 0.884 & 156 & 27 & -126 & 31 & 983 \\
5998.7851 & 0.904 & 144 & 17 & -30 & 41 & 933 \\
5999.5888 & 0.452 & 7 & 18 &  &  & 410 \\
5999.5968 & 0.487 &  &  &  &  & 593 \\
5999.6039 & 0.519 &  &  &  &  & 584 \\
5999.6409 & 0.682 & 151 & 25 & -203 & 20 & 1080 \\
5999.6481 & 0.714 & 141 & 23 & -218 & 19 & 1110 \\
5999.6553 & 0.746 & 167 & 27 & -219 & 20 & 1143 \\
5999.6945 & 0.919 &  &  &  &  & 879 \\
5999.7033 & 0.958 &  &  &  &  & 761 \\
5999.7105 & 0.989 & -8 & 9 &  &  & 680 \\
5999.7439 & 0.137 & -125 & 26 & 165 & 27 & 903 \\
5999.7512 & 0.169 & -142 & 21 & 219 & 23 & 937 \\
5999.7583 & 0.200 & -150 & 25 & 229 & 26 & 957 \\
\hline
\end{tabular}
\end{table*}

36 long-slit spectra were obtained for J150822, and 28 for J160156,
with the Intermediate Dispersion Spectrograph (IDS) on the 2.5~m Isaac
Newton Telescope at La Palma in the Canary Islands.  The observations
for the two stars were interspersed with each other, and covered three
consecutive nights (11-13 March 2012), to optimise phase coverage.
Exposures were 300 or 600~s to allow for the short orbital periods
involved, and a wavelength range of $\sim$7915--9040~$\AA$ was chosen,
which covers the \ion{Ca}{II} triplet.  The RED+2 CCD and R1200R
gratings were used, providing a resolution of 0.51~$\AA$ per pixel.
S/N values of $\sim$40--50 were obtained around quadrature for
J150822, and $\sim$30--40 for J160156.  The spectra were flat-fielded,
bias-corrected and optimally extracted using standard IRAF routines,
and calibrated using CuArNe arc lamp exposures.

Line splitting was clearly observable by eye (Figs.~\ref{linesplit10}
and \ref{linesplit28}) and could be used to estimate phase.  A
suitable synthetic comparison spectrum was then selected by
cross-correlation with a phase 0 program spectrum; the best-matching
template for both objects had a temperature of 4500~K.  Radial
velocities were measured by cross-correlation with the template, using
the IRAF task FXCOR.  Uncertainties were minimized by excluding the
broadest \ion{Ca}{II} line from consideration.  Improved phase
determinations were obtained by fitting sinusoidal functions to the RV
curves to locate cross-over points corresponding to phases 0 and 0.5.

These phasings were then compared with the predictions of the
SuperWASP linear and quadratic ephemerides (implying constant
periods and secular period change respectively), and with internal
simultaneous low-resolution light curves extracted from the spectra
themselves (by evaluating a continuum fit at 8500~$\AA$), and were
found to be substantially self-consistent (Figs.~\ref{star10phasedet}
and \ref{star28phasedet}).  The divergence between different phase
measures is greater for J160156; this is a consequence of its shorter
period, more rapid predicted period change, smaller data sets (both
photometric and spectroscopic) and longer time gap between the last
archived photometry and the spectroscopy.  It is notable that in each
system the deeper minimum of the light curve corresponds to the
eclipse of the secondary, less massive binary component; the SuperWASP
light curves were consequently refolded to locate phase 0 at the time
of true primary eclipse.  The resulting spectroscopic observations and
derived quantities are given in Tables~\ref{star10table} and
\ref{star28table}.  The velocity uncertainties are those obtained with
FXCOR; uncertainties in phase (using the sinusoidal fitting described
above) are negligible in comparison, and were not included in
subsequent modelling.

\section{Results}

\begin{figure}
\resizebox{\hsize}{!}{\includegraphics{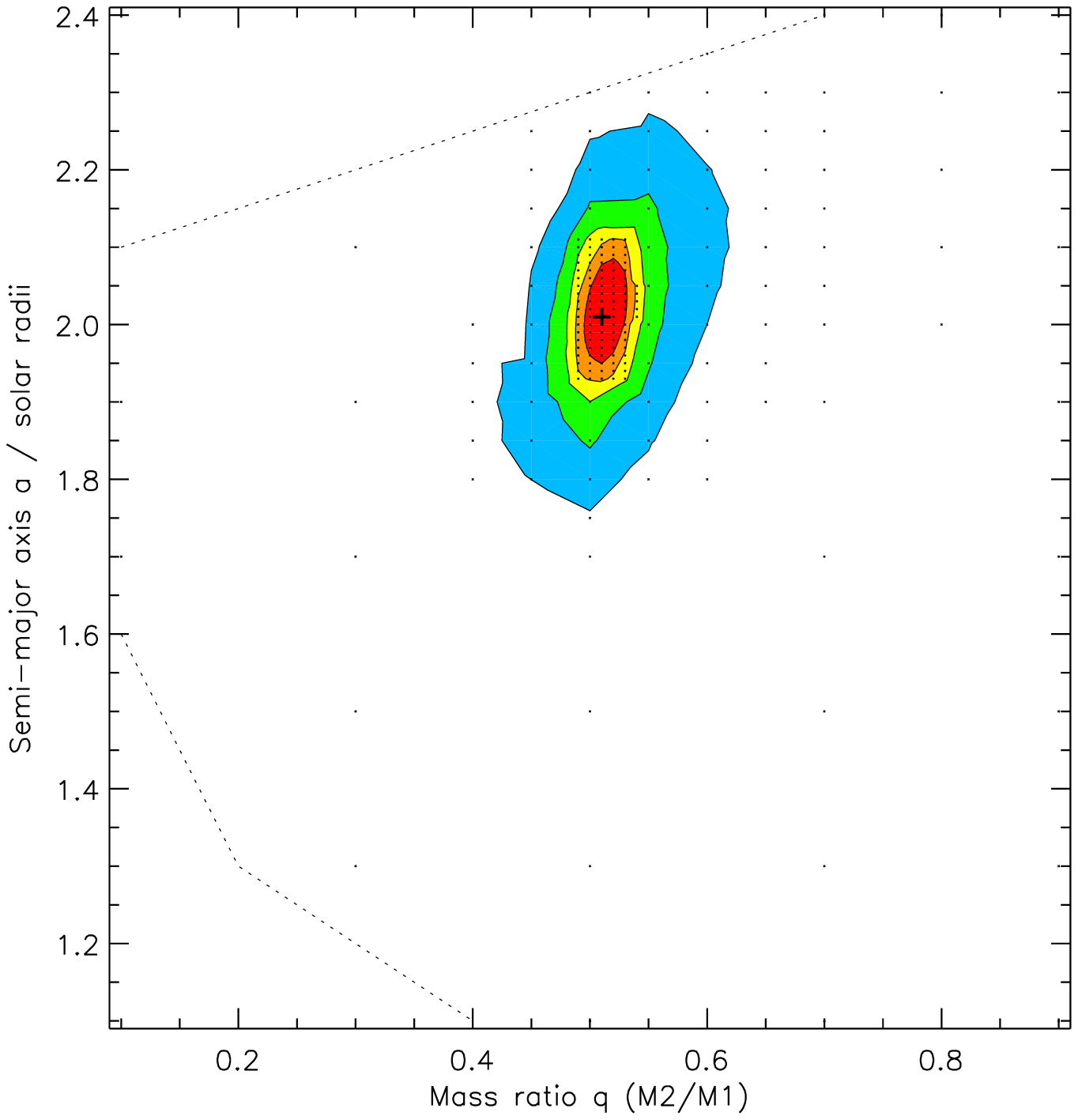}}
\caption{$a$--$q$ parameter cross-section for J150822.  Boxes indicate
points sampled (other parameters being optimized) and the global
minimum is marked with a cross.  Contour lines show the 1, 2, 3, 4 and
5~$\sigma$ uncertainty levels, derived from the $\Delta\chi^2$ values
of the sampled points.  Points outside the plot boundaries or the
dotted lines were not sampled since they corresponded to
physically-implausible masses for the stellar components ($<$0.08 or
$>$1.5~$M_{\sun}$).}
\label{star10contouraq}
\end{figure}

\begin{figure}
\resizebox{\hsize}{!}{\includegraphics{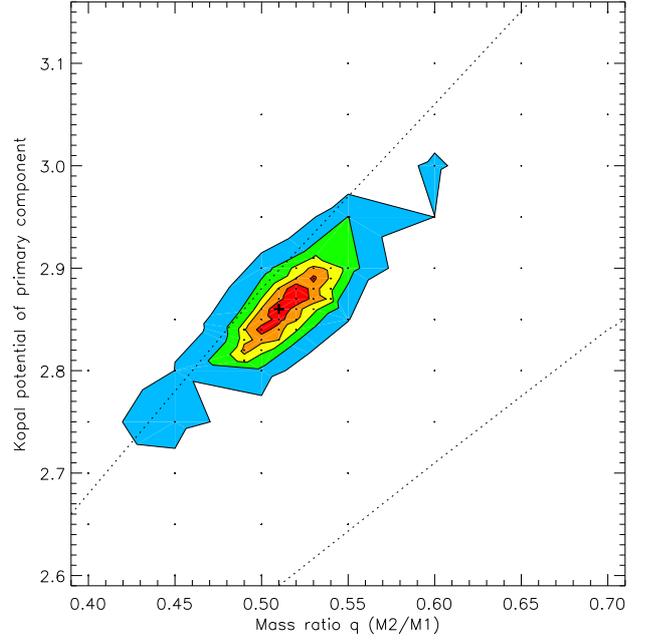}}
\caption{$\Omega_1$--$q$ parameter cross-section for J150822.  Points
below the lower dotted line were not sampled since they corresponded
to physically-implausible filling factors ($F>1$); very high
potentials, corresponding to highly unlikely detached configurations,
were also not sampled.  The upper dotted line indicates the location
of the binary's Roche lobe; the primary component is (with high
probability) just below this line, and hence is likely to be just
overfilling the Roche lobe.}
\label{star10contouro1q}
\end{figure}

\begin{figure}
\resizebox{\hsize}{!}{\includegraphics{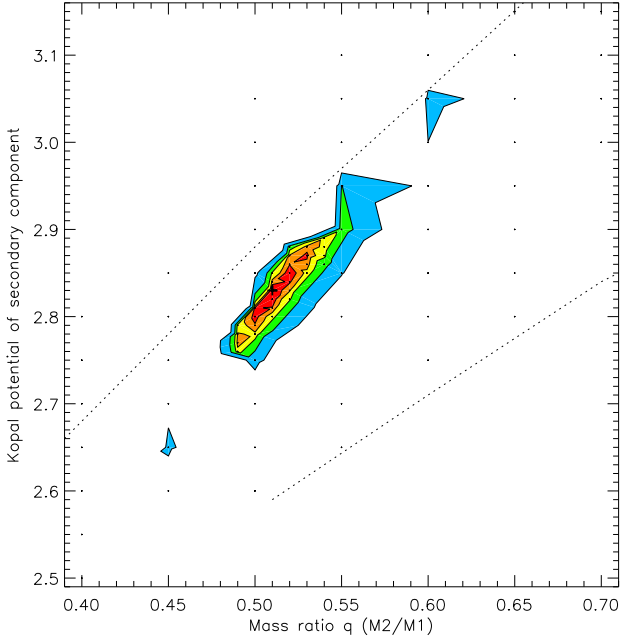}}
\caption{$\Omega_2$--$q$ parameter cross-section for J150822 (see
caption to Fig.~\ref{star10contouro1q} for explanation of dotted
lines).  The secondary component is, with very high probability,
overfilling the Roche lobe.}
\label{star10contouro2q}
\end{figure}

\begin{figure}
\resizebox{\hsize}{!}{\includegraphics{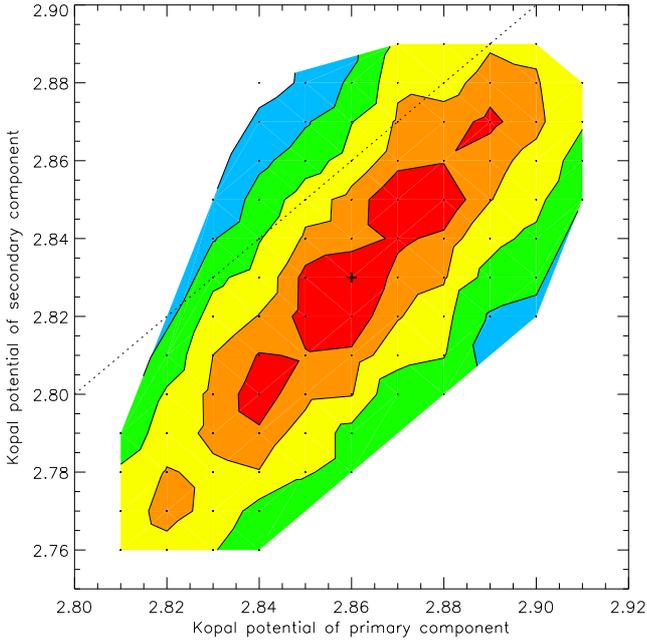}}
\caption{$\Omega_1$--$\Omega_2$ parameter cross-section for J150822.
The dotted line here indicates equal potentials for the two
components, which would necessarily be the case in a contact system;
their probability distribution nearly follows this line, showing a
strong correlation between $\Omega_1$ and $\Omega_2$.}
\label{star10contouromegas}
\end{figure}

\begin{figure}
\resizebox{\hsize}{!}{\includegraphics{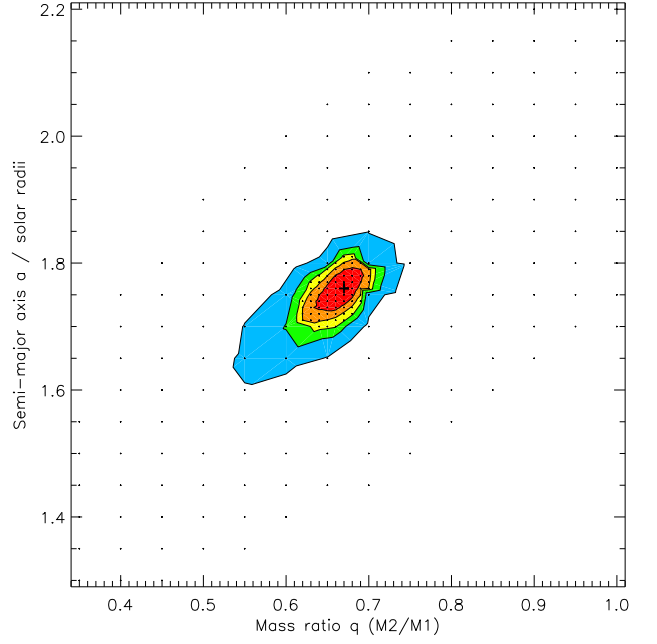}}
\caption{$a$--$q$ parameter cross-section for J160156.  A fairly
strong correlation between semi-major axis and mass ratio is
apparent.}
\label{star28contouraq}
\end{figure}

\begin{figure}
\resizebox{\hsize}{!}{\includegraphics{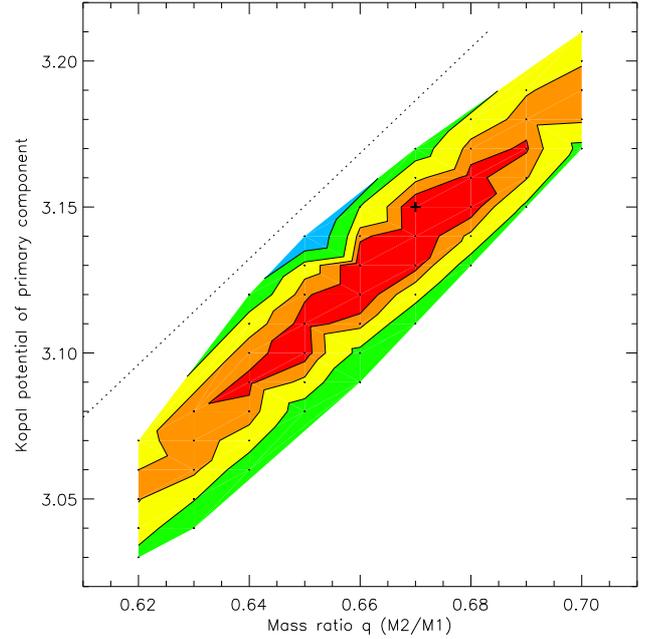}}
\caption{$\Omega_1$--$q$ parameter cross-section for J160156.  The
dotted line indicates the location of the binary's Roche lobe; the
primary component is with high probability below this line, and hence
is very likely to be overfilling the Roche lobe.}
\label{star28contouro1q}
\end{figure}

\begin{figure}
\resizebox{\hsize}{!}{\includegraphics{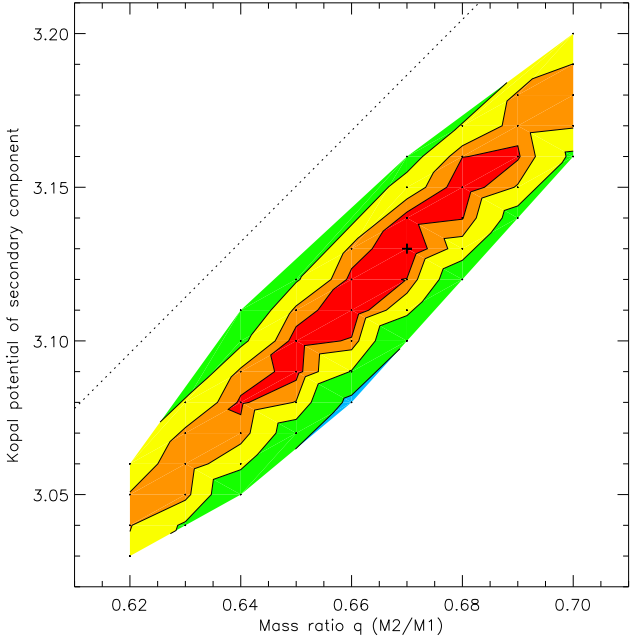}}
\caption{$\Omega_2$--$q$ parameter cross-section for J160156.  The secondary component is also very likely to be overfilling the Roche lobe.}
\label{star28contouro2q}
\end{figure}

\begin{figure}
\resizebox{\hsize}{!}{\includegraphics{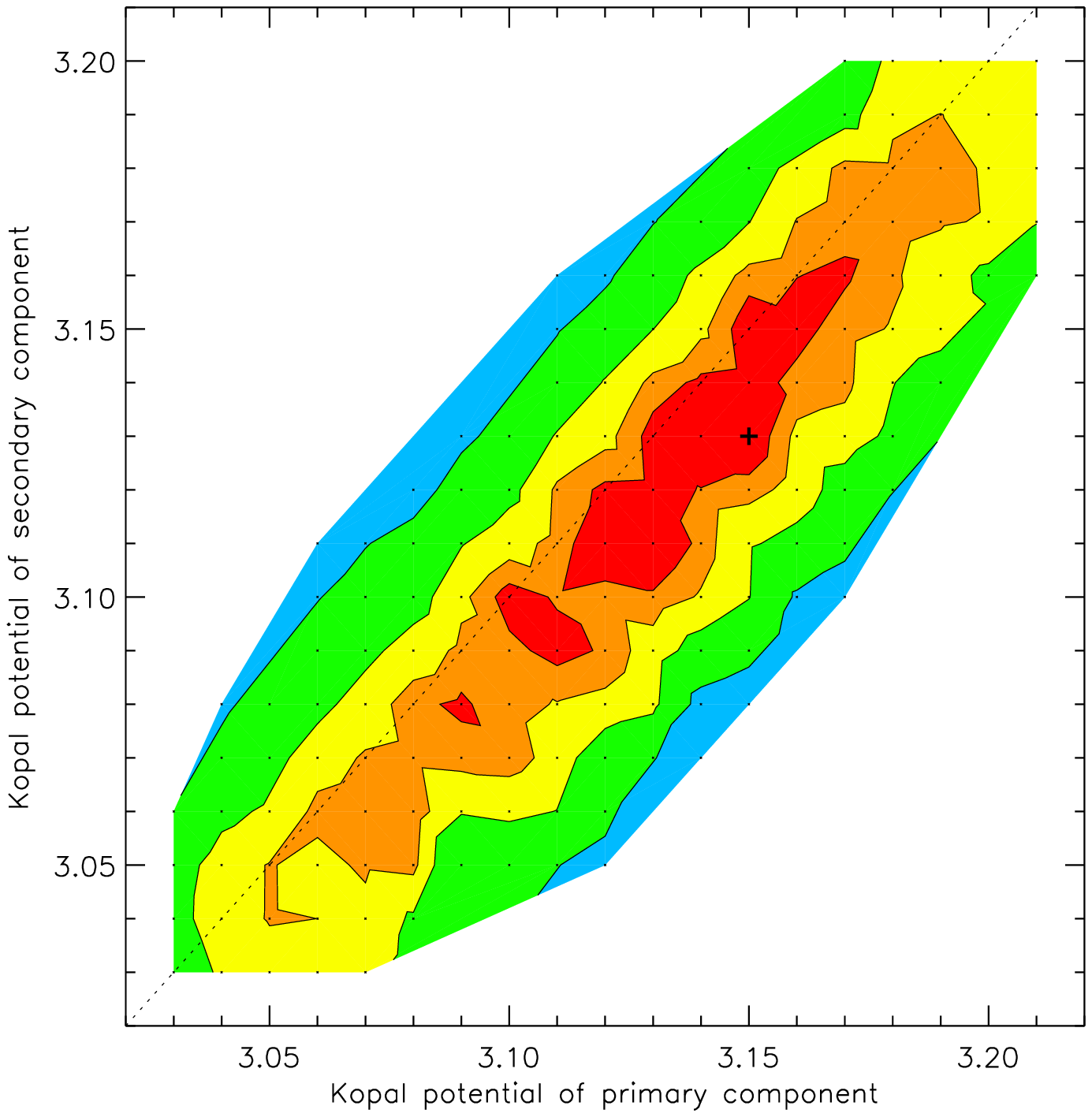}}
\caption{$\Omega_1$--$\Omega_2$ parameter cross-section for J160156.
The region of highest probability closely follows the dotted line, suggesting near-identical potentials for the two components.}
\label{star28contouromegas}
\end{figure}

The eclipsing binary modelling software PHOEBE \citep{prsa}, built
upon the code of \citet{wildev}, was used to model simultaneously the
binned SuperWASP light curves and INT radial velocity curves of the
two systems.  (The full SuperWASP light curves were also
modelled as a final check on the validity of the optima found using
binned curves; it would have been prohibitively time-consuming to
carry out the full modelling procedure using curves consisting of tens
of thousands of data points.) A semi-detached or contact
configuration (one or both components overfilling their Roche
lobes) was assumed on the basis of the continuous light variation in
the light curves, so the Unconstrained mode was used to allow
for both possibilities.  An approximate shared temperature
(which would correspond to the envelope surrounding the
components of a W\,UMa-type system) of $4500\pm250$~K was used for
both EBs, because a template with this effective temperature had
provided the best match for phase 0 and phase 0.5 spectra during
radial velocity determination; this was not varied during modelling
due to the relatively low S/N of the spectroscopic data and the
limited contribution of temperature to the goodness of model fit.

The shortness of the orbital periods involved constrained us to
sub-solar or approximately solar parameters for masses and radii:
large stars simply would not fit into the orbits implied, and so trial
values of semi-major axis were limited accordingly.  No third light
was included since in each light curve the deeper eclipse has roughly
half the flux of the higher maximum.  The details of light curve shape
also constrained the possible angles of inclination: J150822 has
slightly flattened eclipse bottoms, implying $i$ close to 90$\degr$,
while J160156 has more pointed eclipse bottoms, ruling out such a high
angle.  The shapes of the `shoulders' of the maxima in each case
implied Kopal potentials lower than the critical potential at Lagrange
point $L_1$ i.e. yielding binary filling factors in (0,1], using
Pr\v{s}a's definition

\begin{equation}F=\frac{\Omega-\Omega_{\mathrm{crit}}^{L_1}}{\Omega_{\mathrm{crit}}^{L_2}-\Omega_{\mathrm{crit}}^{L_1}}.\end{equation}

The radial velocity curve amplitudes alone determined the semi-major
axes of the orbits $a$ and hence the absolute sizes of the components,
while their mass ratios $q$ were constrained by both light and radial
velocity curves, via the relative eclipse depths and relative
amplitudes of primary and secondary components.  The light curves
provided most of the information needed to determine the optimum
angles of inclination $i$ and Kopal potentials $\Omega_{1,2}$.

Using these guidelines, and following a similar approach to that of
\citet{chew}, an initial best-fit solution was found manually for each
system, which minimized the combined $\chi^2$ values for the light
curve and the two radial velocity curves.  To ensure that these
solutions corresponded to global rather than local minima, to explore
the correlations between fitting parameters, and to determine
realistic uncertainties for the best-fit parameter values, a series of
heuristic scans of the five-dimensional parameter space ($a$, $q$, $i$
and $\Omega_{1,2}$) were carried out using the PHOEBE scripter.
Initially the entire physically-plausible parameter space was scanned
with widely-spaced grids, to ensure that no regions of low $\chi^2$
values had been missed.  The scans were then repeated with decreasing
grid spacings, focusing on regions where the difference
$\Delta\chi^2=\chi^2-\chi^2_{min}$ corresponded to an uncertainty
below 3~$\sigma$ \citep{press}, until the position of the
minimum was determined with accuracy.

The global optima found for the two systems via the scans were
very close to those found manually; the combined minimum $\chi^2$
value for J150822 was 2.36 and for J160156 was 4.72.  Since these
values were far greater than 1, indicating poor model fits (for
reasons explored below), the $\Delta\chi^2$ value at which to set the
1~$\sigma$ uncertainty boundary was set with the assistance of a
separate series of manually-determined optimal solutions for simulated
data sets, with data points randomly perturbed according to their
original individual uncertainties.  The standard deviations of the
parameters, estimated by this method, were comparable in size to the
formal uncertainties given by the Wilson-Devinney covariance matrix.

Figs.~\ref{star10contouraq}--\ref{star28contouromegas}
illustrate some of the 10 two-dimensional parameter cross-sections
obtained from the scans.  The $\Omega_{1,2}$--$q$ planes are
particularly revealing: both systems have best-fit solutions in which
both components overfill their Roche lobes, and where the two
potentials are strongly correlated with each other
(Figs.~\ref{star10contouromegas} and \ref{star28contouromegas}),
suggesting the components of each system share a common potential
within the shared envelope of a contact binary.

\begin{figure}
\resizebox{\hsize}{!}{\includegraphics{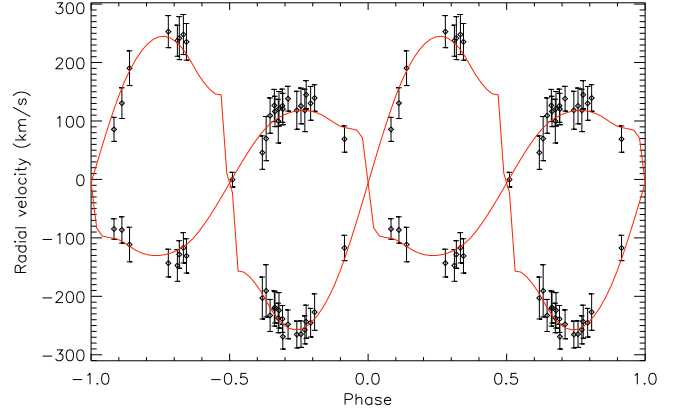}}
\caption{Radial velocity curves for J150822 with best-fit model
overplotted.}
\label{star10rvfit}
\end{figure}

\begin{figure}
\resizebox{\hsize}{!}{\includegraphics{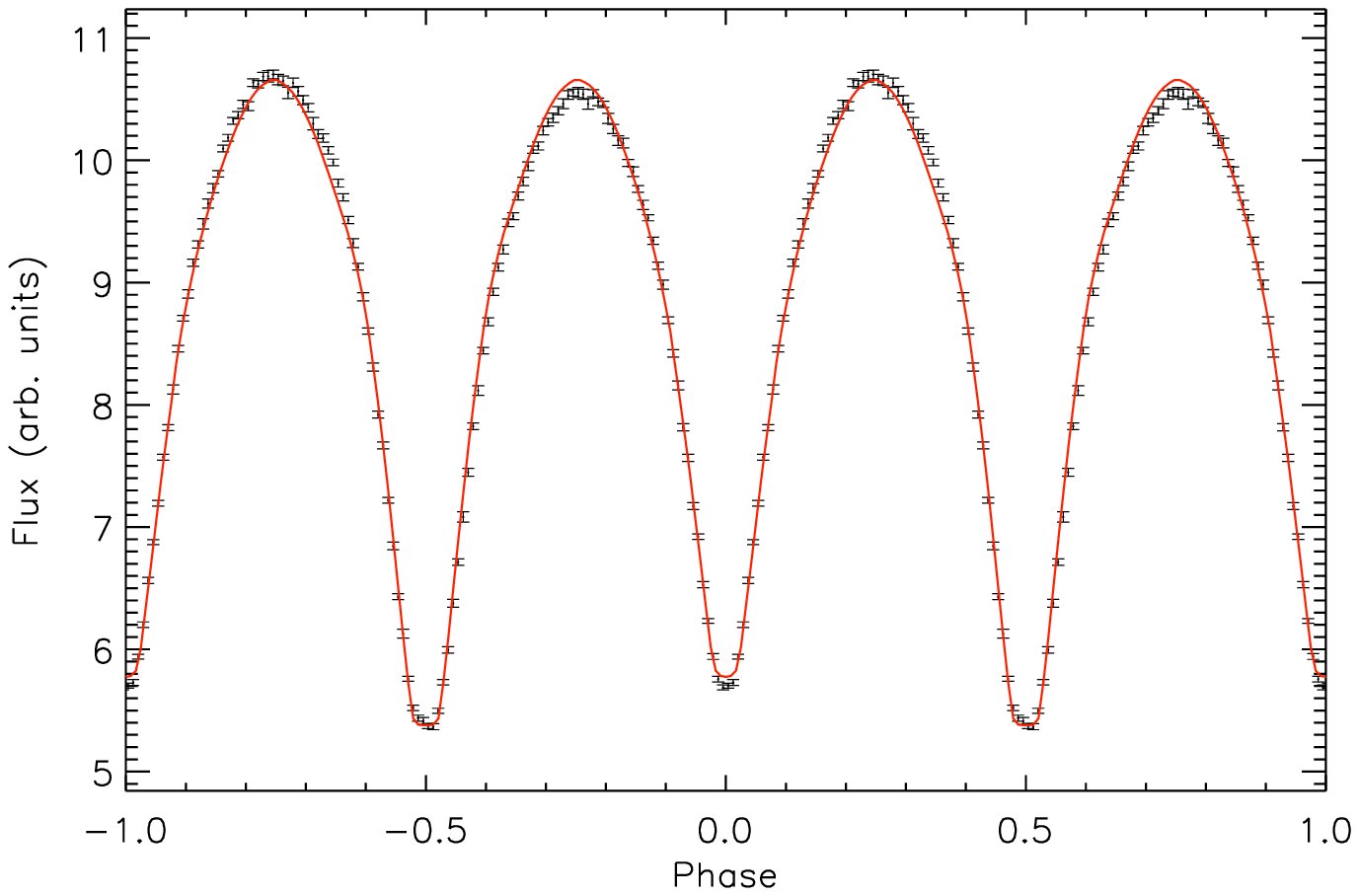}}
\caption{SuperWASP binned light curve for J150822 with best-fit
unspotted model overplotted.}
\label{star10lcfit}
\end{figure}

\begin{figure}
\resizebox{\hsize}{!}{\includegraphics{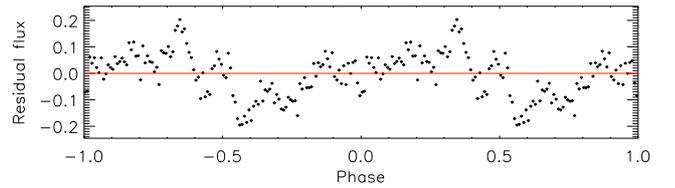}}
\caption{Light curve residuals for J150822 best-fit model.}
\label{star10lcresid}
\end{figure}

\begin{figure}
\resizebox{\hsize}{!}{\includegraphics{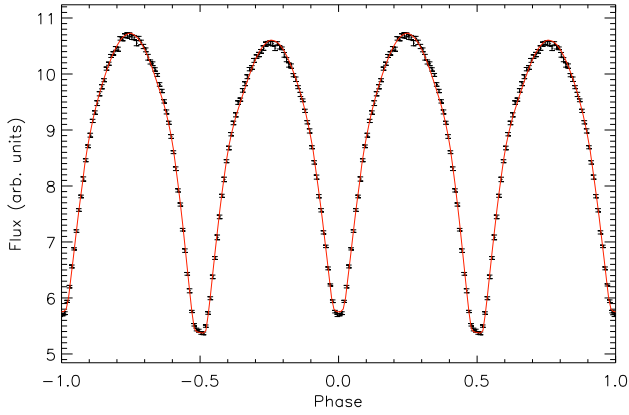}}
\caption{Best-fit model for J150822 with example spot included.}
\label{star10lcspotfit}
\end{figure}

\begin{figure}
\resizebox{\hsize}{!}{\includegraphics{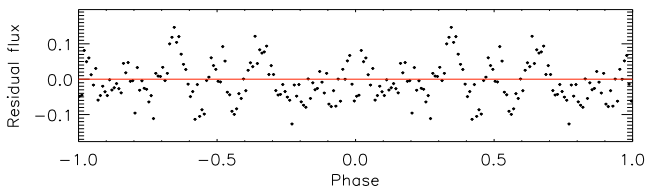}}
\caption{Light curve residuals for J150822 example model with single cool spot.}
\label{star10lcspotresid}
\end{figure}

\begin{figure}
\resizebox{\hsize}{!}{\includegraphics{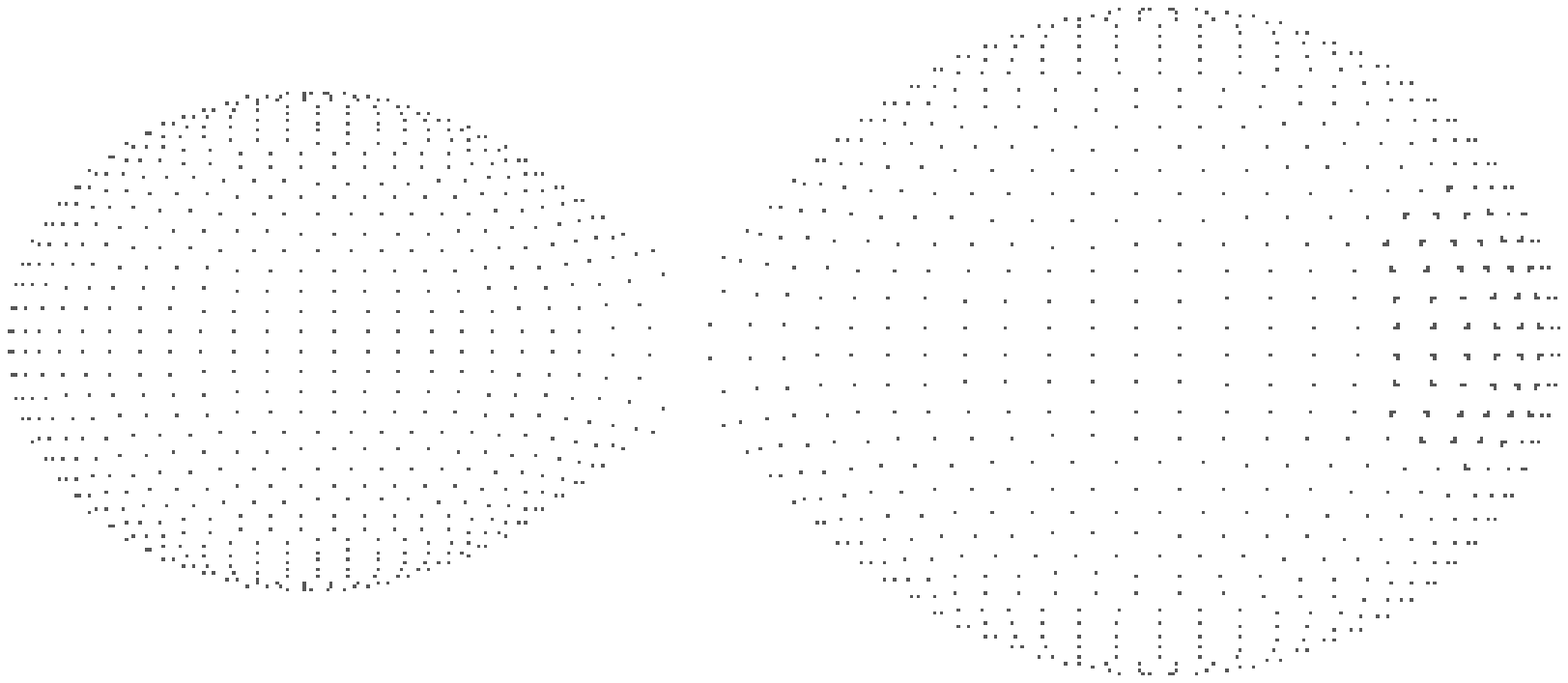}}
\caption{Image of J150822 PHOEBE best-fit model at phase 0.75, indicating
location and size of example cool spot on primary.}
\label{star10spot}
\end{figure}

\begin{figure}
\resizebox{\hsize}{!}{\includegraphics{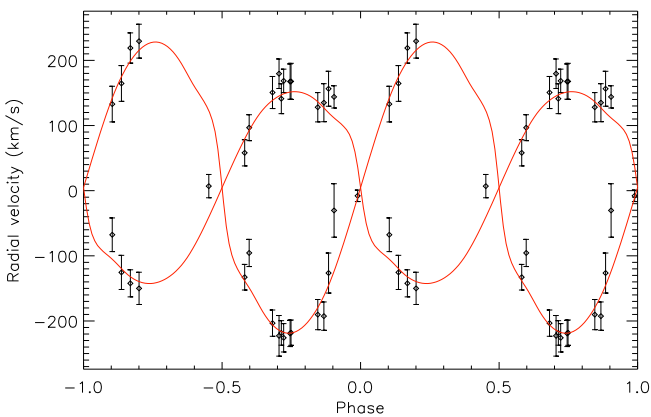}}
\caption{Radial velocity curves for J160156 with best-fit model
overplotted.}
\label{star28rvfit}
\end{figure}

\begin{figure}
\resizebox{\hsize}{!}{\includegraphics{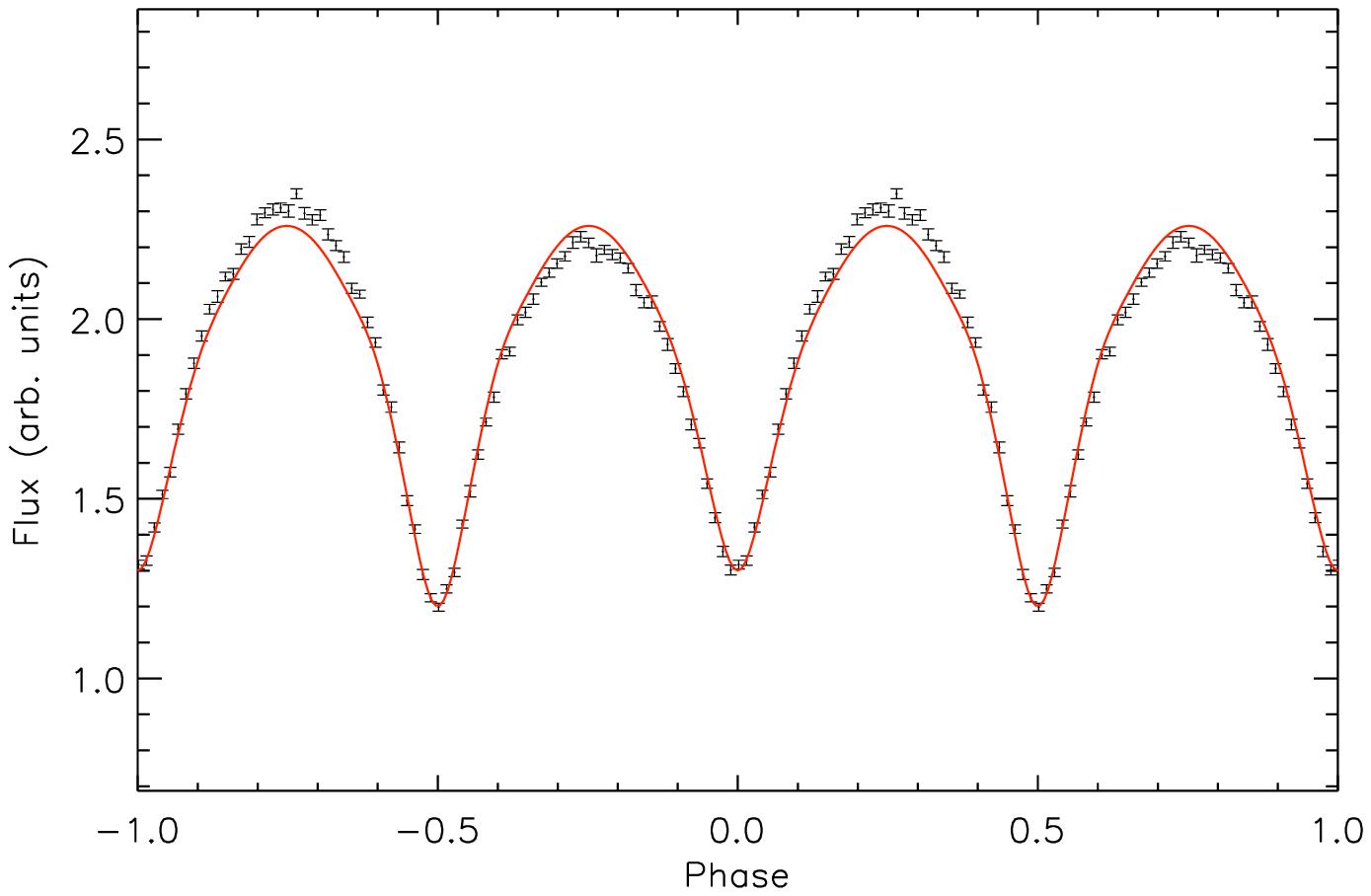}}
\caption{SuperWASP binned light curve for J160156 with best-fit
unspotted model overplotted.}
\label{star28lcfit}
\end{figure}

\begin{figure}
\resizebox{\hsize}{!}{\includegraphics{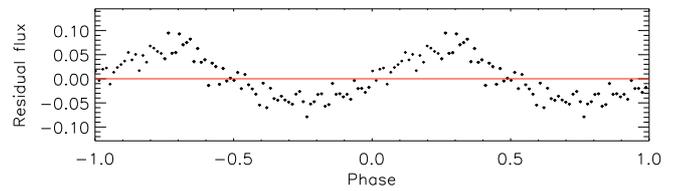}}
\caption{Light curve residuals for J160156 best-fit model.}
\label{star28lcresid}
\end{figure}

\begin{figure}
\resizebox{\hsize}{!}{\includegraphics{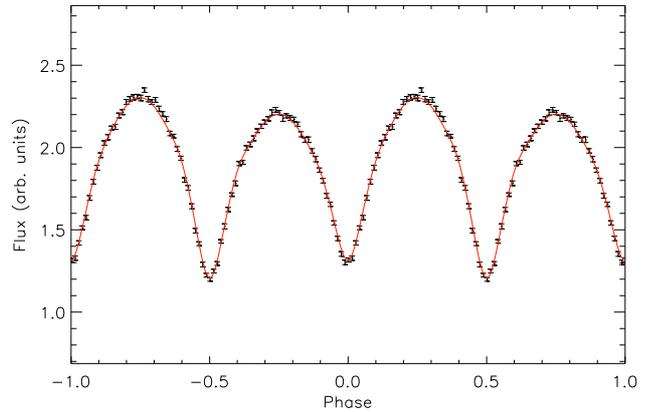}}
\caption{Best-fit model for J160156 with example spot included.}
\label{star28lcspotfit}
\end{figure}

\begin{figure}
\resizebox{\hsize}{!}{\includegraphics{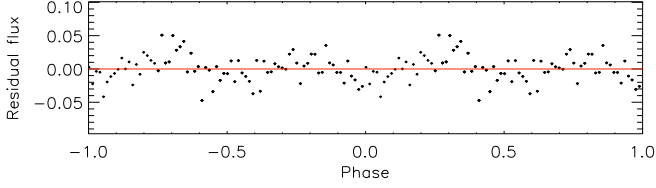}}
\caption{Light curve residuals for J160156 example model with single cool spot.}
\label{star28lcspotresid}
\end{figure}

\begin{figure}
\resizebox{\hsize}{!}{\includegraphics{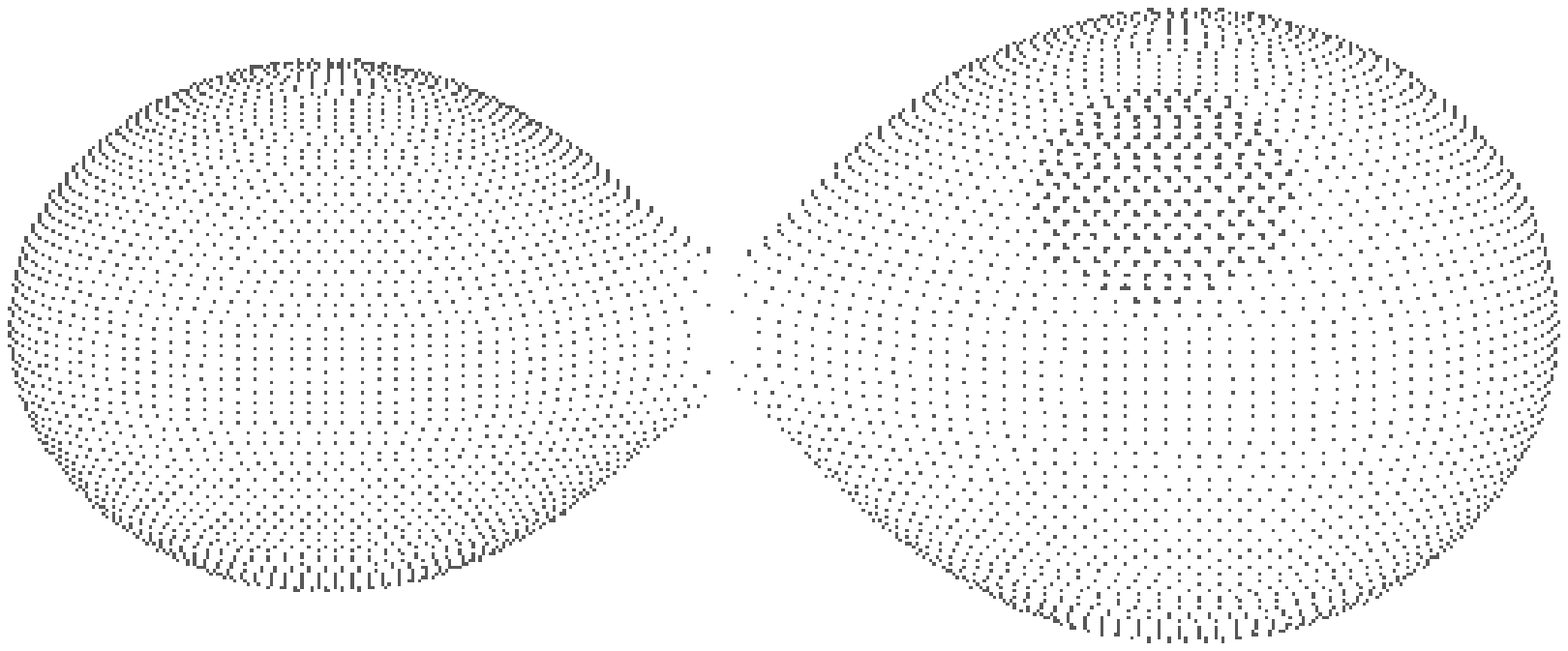}}
\caption{Image of J160156 PHOEBE best-fit model at phase 0.75, indicating
location and size of example cool spot on primary.}
\label{star28spot}
\end{figure}

Figs.~\ref{star10rvfit} to \ref{star28spot} show the best-fit
PHOEBE models for the two systems.  A small Rossiter-McLaughlin
effect \citep{rossiter, mclaughlin} is seen in the model for J150822
in the asymmetry of the radial velocity curves
(Fig.~\ref{star10rvfit}); this is a consequence of the high angle of
inclination.  The different heights of maxima visible in both light
curves, but most notably in J160156, are most likely attributable to
star spots i.e. the O'Connell effect \citep{oconnell}.  However, since
we lack any direct evidence for the number, size or location of spots
(e.g. via Doppler tomography), and being mindful of the additional
modelling latitude provided by inclusion of spots, we sought to
determine the best-fit model for the light curves without including
any spots, and our final stellar parameters result from this model.
Figs.~\ref{star10lcspotfit} and \ref{star28lcspotfit} indicate the
improvement of fit resulting from adding a single cool spot
to the primary component in each system, without altering any
other input parameters.  Figs.~\ref{star10spot} and \ref{star28spot}
show the appearance of the modelled spotted systems, which reproduce
both the different heights of maxima and the different depths of
minima better than the unspotted models can.  Addition of further
spots could doubtless produce a perfect match of models to light
curves, but at the expense of the plausibility of the modelling.

The residuals for the light curve fits reflect these assumed spots in
their large-scale sinusoidal deviations (Figs.~\ref{star10lcresid} and
\ref{star28lcresid}).  However, there are additional clear sinusoidal
variations at a smaller scale in the residuals for J150822
(Fig.~\ref{star10lcspotresid}): an oscillation with an amplitude of
around $\pm$0.1 flux units and a frequency of six cycles per orbit.
Presumably these correspond to pulsation of the primary (since they
are obscured during primary eclipse), and are locked to the binary
orbital period (since they are clearly visible in the folded light
curve).

The final best-fit parameters for both systems are given in
Table~\ref{starparams}.  We would emphasise that these parameters are
not dependent on the inclusion of spots in the models.  The
uncertainties on $a$, $q$, $i$ and $\Omega_{1,2}$ were obtained from
the 1~$\sigma$ contours in the relevant parameter cross-sections.  The
uncertainties on the output parameters (masses and radii) are the
maximum/minimum values obtainable using parameter combinations falling
within these 1~$\sigma$ contours.

\begin{table}
\caption{Modelled system and stellar component parameters for J150822 and J160156}
\label{starparams}
\centering
\begin{tabular}{l l l l}
\hline\hline
 & & J150822 & J160156 \\
\hline
Semi-major axis ($R_{\sun}$) & $a$ & $2.01_{-0.06}^{+0.07}$ & $1.76\pm0.03$ \\
Mass ratio & $M_2/M_1$ & $0.51_{-0.01}^{+0.02}$ & $0.67_{-0.03}^{+0.02}$ \\
COM velocity (km~s\textsuperscript{-1}) & $V_0$ & $-6.2_{-2.5}^{+2.8}$ & $4.7_{-1.2}^{+1.8}$ \\
Angle of incl. ($\degr$) & $i$ & $90_{-3}^{+0}$ & $79.5\pm0.25$ \\
Kopal potentials & $\Omega_1$ & $2.86_{-0.02}^{+0.03}$ & $3.15_{-0.06}^{+0.02}$ \\
 & $\Omega_2$ & $2.83\pm0.04$ & $3.13_{-0.05}^{+0.03}$ \\
Filling factor & $\mathcal{F}$ & $0.12_{-0.04}^{+0.06}$ & $0.10_{-0.00}^{+0.06}$ \\
Masses ($M_{\sun}$) & $M_1$ & $1.07_{-0.09}^{+0.12}$ & $0.86\pm0.04$ \\
 & $M_2$ & $0.55_{-0.05}^{+0.06}$ & $0.57\pm0.04$ \\
Radii ($R_{\sun}$) & $R_1$ & $0.90_{-0.03}^{+0.04}$ & $0.75\pm0.01$ \\
 & $R_2$ & $0.68\pm0.03$ & $0.63\pm0.02$ \\
\hline
\end{tabular}
\end{table}

\section{Discussion}

These results confirm J150822 and J160156, initially identified as
candidate EBs on the basis of light curve shapes alone, as
double-lined spectroscopic and eclipsing binaries.  From modelling,
both systems appear to be composed of late G--early M class dwarfs.
J150822 is slightly more massive, with an approximately solar-mass
primary and late K secondary; its masses have been determined
with a precision of $\sim$10\% and its radii within $\sim$4\%.
J160156's components are of more similar mass: a late G or early K
primary and a late K or early M secondary; its masses have
been found with a precision of $\sim$5\% and its radii within
$\sim$2\%.  Both appear to be W-type systems, in the sense of
\citet{binnendijk}, in that the less massive component is eclipsed
during the deeper minimum.

\begin{figure}
\resizebox{\hsize}{!}{\includegraphics{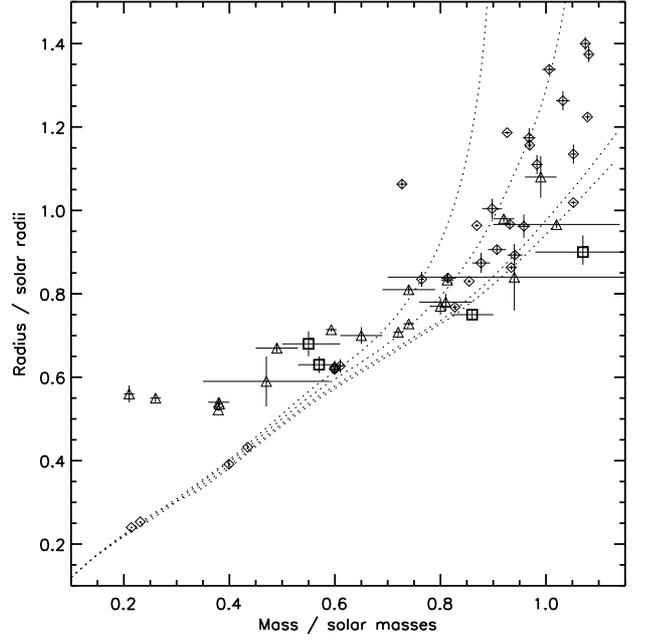}}
\caption{Masses vs. radii of J150822 and J160156 components (squares)
compared with 32 components of low-mass detached binaries (diamonds)
from \citet{torres}, and 20 components of short-period contact
binaries (triangles) from \citet{stepien12}.  Also plotted are
theoretical isochrones derived from Dartmouth models \citep{dotter},
for solar metallicity, with ages of 0.25, 1, 5 and 10~Gyr respectively
(dotted lines, ascending).}
\label{massradplot}
\end{figure}

The contact configuration and likely mass exchange associated
with the apparent period changes make it difficult to compare directly
the mass-radius relationships of these systems with those collected
and discussed in e.g. \citet{torres} (Fig.~2) and \citet{torres13}
(Fig.~4) for detached binaries containing low-mass components.
Fig.~\ref{massradplot} therefore compares our results both with those
of \citeauthor{torres} and of a selection of short-period contact
systems collected by \citet{stepien12}.  We may note that, like
several other contact systems, the binaries studied here are somewhat
discrepant with the Dartmouth model isochrones for solar metallicity
\citep{dotter}\footnote{http://stellar.dartmouth.edu/$\sim$}models/.
Specifically, the primaries have smaller radii than their masses might
suggest, while the secondaries have larger radii than expected.
Possibly the primaries' less dense outer layers have been partially
stripped and transferred to the secondaries, leaving denser
``cores''.  Higher resolution spectroscopy and/or Doppler
tomography would be required for confirmation.

\section{Conclusions}

J150822 and J160156 are established to be spectroscopic double-lined
and eclipsing binary systems in contact configuration, composed of
low-mass dwarfs.  J150822 has been modelled as consisting of 1.07 and
0.55~$M_{\sun}$ components (mass ratio 0.51), and J160156 as having
0.86 and 0.57~$M_{\sun}$ components (mass ratio 0.67).  The primary of
J150822 appears to be pulsating with a period 1/6 of the orbital
period.  Both systems are plausibly undergoing mass transfer; this may
be related to the primaries' radii being smaller, and the secondaries'
radii being larger, than would be typical for single stars with these
masses.

The parameters obtained here should contribute to our understanding of
low-mass stars and contact binary systems, since relatively few
binaries are known with such short orbital periods.  We hope to follow
up further candidate short-period EBs listed in \citet{lohr13} with
multi-colour photometry and spectroscopy, with a view to confirming
their binary nature.  Many of them should be good prospects for full
solution, and capable of significantly extending our knowledge of this
aspect of the field.

\begin{acknowledgements}
The WASP project is funded and operated by Queen's University Belfast,
the Universities of Keele, St. Andrews and Leicester, the Open
University, the Isaac Newton Group, the Instituto de Astrofisica de
Canarias, the South African Astronomical Observatory and by STFC.  The
Isaac Newton Telescope is operated on the island of La Palma by the
Isaac Newton Group in the Spanish Observatorio del Roque de los
Muchachos of the Instituto de Astrofisica de Canarias.  This work was
supported by the Science and Technology Funding Council and the Open
University.  We would like to thank the referee for constructive
comments and recommendations which have improved this paper.
\end{acknowledgements}

\bibliographystyle{aa}
\bibliography{reflist}

\begin{thebibliography}{24}
\expandafter\ifx\csname natexlab\endcsname\relax\def\natexlab#1{#1}\fi

\bibitem[{Binnendijk(1970)}]{binnendijk}
Binnendijk, L. 1970, Vistas in Astronomy, 12, 217

\bibitem[{Chew(2010)}]{chew}
Chew, Y. G.~M. 2010, On the analysis of two low-mass, eclipsing binary systems
  in the young Orion Nebula cluster (unpublished PhD thesis: Vanderbilt
  University)

\bibitem[{Dotter {et~al.}(2008)Dotter, Chaboyer, Jevremovi\'{c}, Kostov, Baron,
  \& Ferguson}]{dotter}
Dotter, A., Chaboyer, B., Jevremovi\'{c}, D., {et~al.} 2008, ApJS, 178, 89

\bibitem[{Jiang {et~al.}(2012)Jiang, Han, Ge, Yang, \& Li}]{jiang}
Jiang, D., Han, Z., Ge, H., Yang, L., \& Li, L. 2012, MNRAS, 421, 2769

\bibitem[{Lohr {et~al.}(2012)Lohr, Norton, Kolb, Anderson, Faedi, \&
  West}]{lohr}
Lohr, M.~E., Norton, A.~J., Kolb, U.~C., {et~al.} 2012, A\&A, 542, A124

\bibitem[{Lohr {et~al.}(2013)Lohr, Norton, Kolb, Maxted, Todd, \&
  West}]{lohr13}
Lohr, M.~E., Norton, A.~J., Kolb, U.~C., {et~al.} 2013, A\&A, 549, A86

\bibitem[{Mazeh {et~al.}(2006)Mazeh, Tamuz, Zucker, Udalski, Consortium,
  Arnold, Bouchy, \& Moutou}]{mazeh}
Mazeh, T., Tamuz, O., Zucker, S., {et~al.} 2006, in Tenth Anniversary of 51
  Peg-b: Status of and prospects for hot Jupiter studies. Colloquium held at
  Observatoire de Haute Provence, France, August 22-25, 2005, ed. L.~Arnold,
  F.~Bouchy, \& C.~Moutou (Paris: Frontier Group), 165--172

\bibitem[{McLaughlin(1924)}]{mclaughlin}
McLaughlin, D.~B. 1924, ApJ, 60, 22

\bibitem[{Norton {et~al.}(2011)Norton, Payne, Evans, West, Wheatley, Anderson,
  Barros, Butters, Cameron, Christian, Enoch, Faedi, Haswell, Hellier, Holmes,
  Horne, Kane, Lister, Maxted, Parley, Pollacco, Simpson, Skillen, Smalley,
  Southworth, \& Street}]{norton}
Norton, A.~J., Payne, S.~G., Evans, T., {et~al.} 2011, A\&A, 528, A90

\bibitem[{O'Connell(1951)}]{oconnell}
O'Connell, D. J.~K. 1951, Publications of the Riverview College Observatory, 2,
  85

\bibitem[{Paczy\'{n}ski {et~al.}(2006)Paczy\'{n}ski, Szczygie\l, Pilecki, \&
  Pojma\'{n}ski}]{paczynski}
Paczy\'{n}ski, B., Szczygie\l, D.~M., Pilecki, B., \& Pojma\'{n}ski, G. 2006,
  MNRAS, 368, 1311

\bibitem[{Pollacco {et~al.}(2006)Pollacco, Skillen, Cameron, Christian,
  Hellier, Irwin, Lister, Street, West, Anderson, Clarkson, Deeg, Enoch, Evans,
  Fitzsimmons, Haswell, Hodgkin, Horne, Kane, Keenan, Maxted, Norton, Osborne,
  Parley, Ryans, Smalley, Wheatley, \& Wilson}]{pollacco}
Pollacco, D.~L., Skillen, I., Cameron, A.~C., {et~al.} 2006, PASP, 118, 1407

\bibitem[{Press {et~al.}(2007)Press, Teukolsky, Vetterling, \&
  Flannery}]{press}
Press, W.~H., Teukolsky, S.~A., Vetterling, W.~T., \& Flannery, B.~P. 2007,
  Numerical Recipes 3rd Edition: The Art of Scientific Computing, 3rd edn. (New
  York: Cambridge University Press)

\bibitem[{Pr\v{s}a \& Zwitter(2005)}]{prsa}
Pr\v{s}a, A. \& Zwitter, T. 2005, ApJ, 628, 426

\bibitem[{Rossiter(1924)}]{rossiter}
Rossiter, R.~A. 1924, ApJ, 60, 15

\bibitem[{Rucinski(1992)}]{ruc92}
Rucinski, S.~M. 1992, AJ, 103, 960

\bibitem[{Rucinski(2007)}]{ruc07}
Rucinski, S.~M. 2007, MNRAS, 382, 393

\bibitem[{Stepie\'{n}(2006)}]{stepien}
Stepie\'{n}, K. 2006, Acta Astron., 56, 347

\bibitem[{Stepie\'{n} \& Gazeas(2012)}]{stepien12}
Stepie\'{n}, K. \& Gazeas, K. 2012, Acta Astron., 62, 153

\bibitem[{Szyma\'{n}ski {et~al.}(2001)Szyma\'{n}ski, Kubiak, \&
  Udalski}]{szymanski}
Szyma\'{n}ski, M., Kubiak, M., \& Udalski, A. 2001, Acta Astron., 51, 259

\bibitem[{Tamuz {et~al.}(2005)Tamuz, Mazeh, \& Zucker}]{tamuz}
Tamuz, O., Mazeh, T., \& Zucker, S. 2005, MNRAS, 356, 1466

\bibitem[{Torres(2013)}]{torres13}
Torres, G. 2013, Astron. Nachr., 334, 4

\bibitem[{Torres {et~al.}(2010)Torres, Andersen, \& Gim\'{e}nez}]{torres}
Torres, G., Andersen, J., \& Gim\'{e}nez, A. 2010, A\&AR, 18, 67

\bibitem[{Wilson \& Devinney(1971)}]{wildev}
Wilson, R.~E. \& Devinney, E.~J. 1971, ApJ, 166, 605

\end{thebibliography}

\end{document}